\documentclass[10pt,amsmath,amssymb,superscriptaddress]{iopart}
\usepackage{iopams}
\usepackage{setstack}
\usepackage[dvips]{graphicx}
\usepackage{epsfig}
\usepackage{dcolumn}
\usepackage{bm}
\usepackage{longtable}

\begin{document}

\title{A Two-populations Ising model on diluted Random Graphs}

\author{E. Agliari$^{1,2,3}$, R. Burioni$^{1,2}$ and P. Sgrignoli$^{1}$}
\address{$^1$ Dipartimento di Fisica, Universit\`a di Parma, Viale
G.P. Usberti n.7/A  43100 - Parma - ITALY}
\address{$^2$ INFN, gruppo collegato di Parma, Viale G.P. Usberti
n.7/A 43100 - Parma - ITALY}
\address{$^3$ Theoretische Polymerphysik, Universit\"{a}t Freiburg,
Hermann-Herder-Str. 3, D-79104 Freiburg, Germany}

\begin{abstract}
We consider the Ising model for two interacting groups of spins embedded in an Erd\"{o}s-R\'{e}nyi random graph. The critical properties of the system are investigated by means of extensive Monte Carlo simulations. Our results evidence the existence of a phase transition at a value of the inter-groups interaction coupling $J_{12}^C$ which depends algebraically on the dilution of the graph and on the relative width of the two populations, as explained by means of scaling arguments. We also measure the critical exponents, which  are consistent with those of the Curie-Weiss model, hence suggesting a wide robustness of the universality class.
\end{abstract}

\section{Introduction}
\label{intro}
Statistical ferromagnetic models on diluted random graphs have received  an increasing attention 
in the last few years. Although quenched randomness has been primarily 
discussed in the framework of antiferromagnetic models and competing interactions \cite{grest,binder,usadel,dede}, a renewed interest has recently emerged for quenched disorder in models with purely ferromagnetic interactions \cite{bovier,agliari1,gulielmo,almeida,montanari,dsg,fanelli,nostro,agliari2,capitolo}. 
Many problems are still open in this field, and the effects
of quenched topology on the critical properties has not yet been exhaustively clarified.  

In $d$-dimensional disordered  and inhomogeneous systems, the relevance of topological disorder at the phase transition can often be estimated by extended Harris criteria \cite{chayes,luck,weigel}, taking into account  the role of geometrical fluctuations. In these cases, the influence of disorder on the critical behavior can be analyzed by estimating the shift  in the critical temperature induced by disorder fluctuations in a region of spin variables, included in a correlated volume $\xi^d$, where $\xi$ is the  correlation length. When the fluctuation grows in approaching the critical point, a crossover phenomenon can occur and the critical behavior changes.  In mean field models, it is reasonable to expect that in high dimensions the critical behavior should not by be influenced by the presence of topological disorder. In this direction, one of the main points reached by recent studies on purely ferromagnetic models is that the mean-field  behavior, coinciding with a complete graph topology, is very robust with respect to the random dilution of  Erd\"{o}s-R\'{e}nyi random graphs \cite{agliari1,barra}. The universality class of the diluted Ising model appears independent of dilution and also topologies of graphs with arbitrary degree distributions, provided the convergence of the second moment, still feature an analogous behavior \cite{Leonezecchina}.  However, a few rigorous results are available in this case \cite{dembo,cristian}, and estimates of critical exponents are often performed via numerical analysis. 

An interesting example of a ferromagnetic model on a random topology is represented by the two-populations Ising model.  On the randomly diluted graph, two types of spins are present, with an interaction that is homogeneous among the spins belonging to each group, while it is characterized by a different value between the two groups.  The possibility of considering two types of  magnetic sites, interacting therefore through a four blocks matrix, appears an interesting point, which could be relevant in many phenomena, ranging from the study of anisotropic magnetic materials \cite{schiffer}  to social economic models \cite{contucci1}, to biological systems \cite{mello}. In the latter case, an interesting application
concerns cooperativity effects in the context of bacterial chemotaxis, where cooperativity between receptors has been shown to be a key mechanism for a long-lasting puzzle, namely how a small change in external concentration of attractant or repellent can cause significant amplification in receptor signal \cite{segall,mello}.  
Theoretical models based on the two-population Ising model, although simplified, are able to show that a coupled system of receptors has the capacity to greatly amplify signals. These results suggest that the interactions among different types of receptors attain optimal value as it approaches the critical coupling energy of an analogous Ising model; the understanding of the pertaining critical properties is therefore very looked for \cite{bray}.

Of course, all the cited applications require an extension of mean-field results to more general topologies. Diluted random graphs represent a first step in this direction.
The mean field version, corresponding to a complete graph, of the two-populations Ising model has recently been
investigated analytically \cite{contucci2}. There, the phase diagram of the model was studied, as a function of the inter-groups interactions $J_{12}$, of the external fields and of the relative width of the  two groups. In this paper we aim at extending these results to more general topologies, by investigating the critical properties of the two-population Ising model on a diluted random graph, as a function of the inter-groups interaction coupling $J_{12}$. First of all, we show that the simpler cases of two-population models on fully-connected  and diluted bipartite topologies, as well as one-population models on Erd\"{o}s-R\'enyi graphs can be recovered by properly tuning the system parameters, i.e $J_{12}$, the relative width of the two groups $c$ and the degree of dilution $\alpha$. Moreover, non-trivial thermodynamic behavior can be achieved when the intra-group couplings are fixed at a small value; then, there exists a critical value $J_{12}^C$ at which the system exhibits a 
mean-field type phase transition, in agreement with the fully connected model. We determine numerically the critical value of the inter-groups interaction coupling $J^C_{12}$, which turns out to depend algebraically on $\alpha$ and on $c$ and we explain this algebraic behavior by means of scaling arguments. We also measure the critical exponents, showing that they are consistent with those of the Curie-Weiss model, hence suggesting a wide robustness of the universality class. 

In the following, we first describe the mathematical details of the model considered (Section \ref{sec:model}), then we present our numerical and analytical results (Section \ref{sec:numerics}).
The last section is devoted to conclusions and final remarks (Section \ref{sec:conclusions}).

\section{The model} \label{sec:model}
Given a set of $N$ spins denoted as $\{ \mathbf{\sigma} \} \equiv \{\sigma_1, \sigma_2,...,\sigma_N \}$, and arranged according to a given topology, we define the Hamiltonian
\begin{equation} \label{eq:Hamiltonian}
H_N(\{\sigma\})=-\frac{1}{2N}\sum_{\begin{array}{c} \scriptstyle{i, j=1} \\  {\scriptstyle i \sim j}\end{array}}^N J_{ij}\sigma_i\sigma_j-\sum_{i=1}^N h_i\sigma_i \; ,
\end{equation}
where the first sum runs over all couples of nearest neighbours $i \sim j$ and the whole set $I$ of spins $\sigma_i$ is divided into two partitions $I_1 \cup I_2 = I$, made up of $N_1=|I_1|$ and $N_2=|I_2|$ elements, being $N_1+N_2=N$. Given two spins $\sigma_i$ and $\sigma_j$, their interaction depends on the relative subset they belong to, as encoded by the coupling block matrix, which reads:

\begin{displaymath}
  \mathbf{J}=
  \begin{array}{ll}
    \quad
    \overbrace{\qquad}^{N_1}
    \overbrace{\qquad \quad}^{N_2}
    \\
    \left(\begin{array}{c|ccc}
        \mathbf{ J_{11}} &  & \mathbf{ J_{12}}
        \\
        \hline
        &  &  &
        \\
        \mathbf{ J_{21}} &  & \mathbf{ J_{22}}
        \\
        &   &   &
        \\
      \end{array}\right)
  \end{array}
  \!\!\!\!\!
  \begin{array}{ll}
    \\
    \left\} \begin{array}{ll}
        \\
      \end{array}  \right. \!\!\!\!\! N_1
    \\
    \left\} \begin{array}{ll}
        \\
        \\
        \\
      \end{array} \right. \!\!\!\!\! N_2
  \end{array}
\end{displaymath}

In the following we consider only ferromagnetic interactions, i.e. $J_{ij} >0$ for any $i$ and $j$; also, $\mathbf{J}$ will be assumed to be symmetric, i.e. $J_{12} = J_{21}$. 
Therefore, the system under study displays three degrees of freedom: $J_{11}$ and $J_{22}$, tuning the interactions within each subgroup, and $J_{12}$ tuning the interaction between spins of different subsets.  

As for the external field $\mathbf{h}$, it assumes two values $h_1$ and $h_2$, according to the subset it is applied on, as described by the following vector:

\begin{displaymath}
  \mathbf{h}=
  \left(\begin{array}{ccc|c}
      h_{1}
      \\
      \hline
      \\
      h_{2}
      \\
      \\
    \end{array}\right)
  \!\!\!
  \begin{array}{ll}
    \left\} \begin{array}{ll}
        \\
      \end{array}  \right. \!\!\!\!\! N_1
    \\
    \left\} \begin{array}{ll}
        \\
        \\
        \\
      \end{array}  \right. \!\!\!\!\! N_2
    \!\!\!\!\!\!
  \end{array}  
\end{displaymath}

We now introduce the magnetization for the subset $I_k$ as
\begin{equation}\label{eq:mag}
  M_k(\{\sigma\}_k)=\frac{1}{| I_k|}\sum_{i \in I_k}\sigma_i,
\end{equation}
and we define $c \equiv N_1/N$ the fraction of spins belonging to $I_1$, it follows straightforward that $N_2 = (1-c)N$. The total magnetization is given by $M (\{ \sigma \}) = M_1 (\{ \sigma \}_1) + M_2 (\{ \sigma \}_2)$.

The system under investigation has been studied analytically in the special case of a complete graph $K_N$ (each node is linked to any other $N-1$ nodes) as underlying structure \cite{contucci1,contucci2}. 
In this case the Hamiltonian per spin can be rewritten as
\begin{eqnarray} \label{eq:contucci}
  \frac{H_N}{N} &=& -\frac{1}{2} \left[ J_{11}c^2M_1^2+2J_{12}c(1-c)M_1M_2+J_{22}(1-c)^2M_2^2 \right]+ \nonumber \\
  && \, -h_1cM_1 - h_2(1-c)M_2,
\end{eqnarray}
where we dropped the obvious dependence on the magnetic configuration $\{ \sigma \}$. Then, assuming that states are distributed according to the Boltzmann-Gibbs measure and in the thermodynamic limit, the expected value of the total magnetization was shown to depend on partial magnetization according to
\begin{equation}
  \left<M\right>=c M_1+(1-c) M_2.
\end{equation}
Moreover, the implicit analytic solution for the model reads as
\begin{equation} \label{eq:solContucci}
  \left\{
    \begin{array}{rl}
      M_1=\tanh [J_{11}cM_1+J_{12}(1-c)M_2-h_1] \nonumber \\
      M_2=\tanh [J_{12}cM_1+J_{22}(1-c)M_2-h_2] ,
    \end{array}
  \right.
\end{equation}
holding in general over the whole parameter space $(c, \mathbf{J},\mathbf{h})$.

Let us now turn to the case of a diluted graph as substrate. More precisely, we considered Erd\"{o}s-R\'enyi random graphs  $\mathcal{G}(N,p)$ made up of $N$ sites where links between sites are drawn independently with probability $p = \alpha/N$, in such a way that the average coordination number, or degree, is given by $\alpha$. In this work we will always choose $\alpha>1$, so that the resulting graph is overpercolated.
In the following we will neglect the external field, fixing $h_1= h_2 = 0$ in order to focus on the role of $\alpha$ and $c$.

\subsection{Bounds of the model}

The Hamiltonian introduced in Eq.~\ref{eq:Hamiltonian} and endowed with a symmetric $\mathbf{J}$ can be rewritten introducing the local fields $\varphi_k^i$, acting on the spin $i$ and depending on the magnetic configuration of the subset $I_k$, with $k=1,2$:
\begin{equation}
\varphi_k^i \equiv \frac{J_{kk}}{2N} \sum_{\begin{array}{c} \scriptstyle{j \in I_2} \\  {\scriptstyle j \sim i}\end{array}} \sigma_j.
\end{equation}
Hence, the Hamiltonian can be rewritten as
\begin{equation} \label{eq:Hamiltonian}
H_N(\{\sigma\})=-\frac{J_{12}}{2N}\sum_{i \in I_1} \sum_{\begin{array}{c} \scriptstyle{j \in I_k} \\  {\scriptstyle j \sim i}\end{array}}  \sigma_i \sigma_j - \sum_{i \in I_1} \varphi_1^i \sigma_i - \sum_{i \in I_2} \varphi_2^i \sigma_i,
\end{equation}
the first term accounts for the interaction between spins belonging to different populations, while the last two terms for the interaction with local fields. As long as the latter terms are negligible with respect to the former, the system approaches a bipartite, diluted Ising model. Conversely, when the latter terms prevail the system approaches two independent, diluted Ising ferromagnets.

By denoting with $\overline{ \cdot }$ the expectation value of an observable with respect to the possible realizations of dilution and recalling $\sigma_j = \pm 1$, we get the following bounds
\begin{equation}
0 \leq | \overline{ \varphi_k^i} | \leq \frac{J_{kk} (N_k -1 )}{ 2N} p,
\end{equation}
where, for $N_k \gg 1$, the upper bounds can be approximated as $J_{11}pc/2$ and $J_{22}p(1-c)/2$, respectively. 

Now, when $p$ approaches $1$, the substrate recovers the complete graph $K_N$ and the self-consistent equations (\ref{eq:solContucci}) hold. Conversely, when $p \ll 1$ (yet over the percolation threshold), assuming $J_{11}/J_{12} \ll 1$ and $J_{22}/J_{12} \ll 1$, as explained in Sec.~\ref{sec:numerics}, we have that if $c$ is close to $1/2$, the system approaches the case of a diluted ferromagnet, with unitary coupling, set at temperature $J_{12}^{-1}$ and with rescaled dilution $\alpha/2$, in such a way that a phase transition is expected to occur at $J_{12}^C = 2/p$ \cite{agliari1}.
On the other hand, if $c \ll 1 $, (and symmetrically if $c$ close to $1$), the smallest subset $I_1$ significantly interacts only with spins belonging to the other group while the largest subset $I_2$ can display comparable interactions with both subgroups, up to very large couplings $J_{12} \gg J_{22}(1/c-1)$, when we recover again the case of a diluted ferromagnet. Therefore, once $J_{11}$ and $J_{22}$ fixed as explained below, it is intuitive to look at $J_{12}$ as an inverse ``temperature'' of the system, tuning the competition between an energetic and an entropic term: by lowering $J_{12}$ a spin-flip on a node connected with both parties gets more and more likely. 

In the following we will focus on non-trivial cases corresponding to small $c$ and small $p$. 

\section{Numerical results}\label{sec:numerics}

The critical behaviour of the system described by the Hamiltonian of Eq.~\ref{eq:Hamiltonian} has been investigated by means of Monte Carlo (MC) simulations \cite{MC} based on the single spin-flip Glauber algorithm according to which, given the magnetic configuration $\{\sigma\}$, the spin-flip $\sigma_j \rightarrow \sigma_j' = - \sigma_j$ on the $j$-th site, extracted randomly, is accepted with probability 
\begin{equation} \label{eq:glauberProb}
p(\{\sigma\},\sigma'_j, \mathbf{J})=\frac{1}{1+e^{\Delta H_N(\{\sigma\},\sigma'_j, \mathbf{J})}} \; ,
\end{equation}
where $\Delta H_N(\{\sigma\},\sigma'_j,\mathbf{J})=H_N(\{\sigma\}, \sigma'_j)-H_N(\{\sigma\}, \sigma_j)$ is the variation in the system energy due to spin-flip. 
As well-known, such a dynamic is able to lead the system towards equilibrium states described by the canonical distribution. More precisely, we checked that, once the coupling matrix $\mathbf{J}$ and the parameters $c$ and $\alpha$ are fixed, the system eventually reaches a stationary state which, for $N$ large enough so to avoid finite size effects, depends on the above mentioned parameters while it is independent of the initial configuration chosen. Moreover, the fluctuations of a specific observable $x\ (\{\sigma\},\alpha,c,\mathbf{J})$ scale like $N^{-\frac{1}{2}}$.

Therefore, the estimate of a given observable $\left<x\right>$ is taken to be the average over a number of $10^3$ decorrelated states of the system, once the equilibrium regime has been reached. The thermalization time and the decorrelation time are taken to be order of $10^2$ MC steps; of course, as the critical region is approached (see below) the decorrelation time gets larger, nevertheless this choice of parameters ensures a good compromise between the number of measurements and the length of the run, allowing a proper sampling of the phase space.
Moreover, the thermal averages $\left<x\right>$ obtained this way are further averaged over different ($\sim 100$) realizations of the underlying structure (with fixed number of nodes, average coordination number and populations width) in order to account for the stochasticity of the graph; however, in general, statistical errors due to ``topological average'' are significantly smaller than those arising from the thermal average.

We focused in particular on the magnetizations $M_1$, $M_2$ (see Eq.~\ref{eq:mag}) and on the susceptibility $\chi$; the latter can be estimated as:
\begin{equation}\label{eq:susc}
\chi = J_{12} [\langle M^2 \rangle - \langle M \rangle^2],
\end{equation}
recalling that here $J_{11}$ and $J_{22}$ are fixed, $J_{12}$ is finite and $J_{12} \gg J_{11},J_{22}$

Before proceeding we notice that we can distinguish three different cases, according to the values of $J_{11}$ and $J_{22}$: we say that $J_{11}$ is small (large) if the isolated, i.e. $J_{12}=0$, subsystem $\{ \sigma_i\}_{i \in I_1}$ is spontaneously in a paramagnetic (ferromagnetic) state; analogously for $J_{22}$. 
Thus, if we fix $J_{11}$ and $J_{22}$ both large, for $J_{12}=0$ we have that $|M_1|=|M_2|=1$, and increasing $J_{12}$ the (absolute) magnetization remains always larger than zero. Analogously, no phase transition is expected for the case $J_{11}$ large ($|M_1|=1$) and $J_{22}$ small ($M_2=0$), or vice versa. In fact, by rising $J_{12}$ from $0$, the total magnetization will increase continuously from $M_1$ to $M_1+M_2$. The effect is analogous to the case of a ferromagnet (system $2$) set at a temperature larger than the critical one and coupled with an external field (system $1$).
 
 In the following analysis we will therefore focus on the most interesting case of small $J_{11}$ and small $J_{22}$.

\subsection{Critical coupling $J_{12}^C$} \label{ssec:coupling}

\begin{figure}[ht]
  \centering
  \includegraphics[width=.85\textwidth]{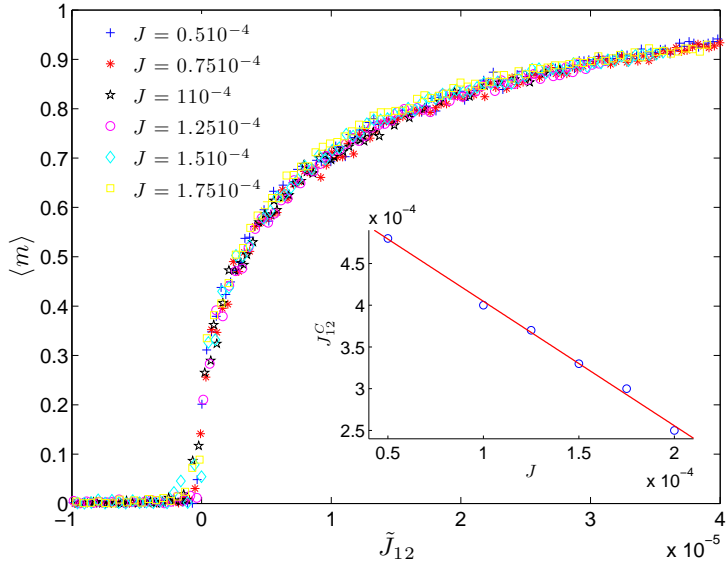}
  \caption{Main figure: Magnetization versus rescaled coupling $\tilde{J}_{12}$ (see Eq.~\ref{eq:scaling}, here the exponent $B=1/3$ was used) for systems of $N=6000$ nodes, with $p=0.80$ and $c=0.20$; several choices of $J_{11}=J_{22}=J$ are considered, as explained in the legend. 
Inset: $J_{12}^C$ as a function of $J$; data points ($o$) are fitted by a linear law (continuous line): in agreement with Eq.~\ref{eq:J_quad}, when $J_{11}=J_{22}$ the critical coupling scale like $\sim - J$.}
  \label{fig:robusto}
\end{figure}

As shown in Fig.~\ref{fig:robusto}, for sufficiently large interaction strength $J_{12}$ the system exhibits a phase transition and spontaneous magnetization occurs even in the absence of an external magnetic field. The value at which the phase transition occurs is denoted $J_{12}^C$ and it is found to depend on the intra-population couplings $J_{11}$ and $J_{22}$, on the graph dilution $\alpha$ as well as on the relative size of the populations $c$.

In order to evaluate the critical coupling and by this the location of the transition, we fitted numerical data found for the magnetization and for the susceptibility using the following relations holding when the critical point is approached: 
\begin{equation}\label{eq:m} 
M \sim \left(1 - \frac{J_{12}^C}{J_{12}} \right)^\beta, \;\;\;\; \mathrm{with} \; J_{12}>J_{12}^C
\end{equation}
and
\begin{equation}\label{eq:chi}
\chi \sim \left|1 - \frac{J_{12}^C}{J_{12}}  \right|^\gamma.
\end{equation}
The measurements of $J_{12}^C$ obtained independently from Eq.~\ref{eq:m} and from Eq.~\ref{eq:chi} 
are, within the error (approximately $3 \%$), consistent with each other.
In the next subsection we will exploit these values to determine the critical exponents $\beta$ and $\gamma$.

\begin{figure}[ht]
  \centering
  \includegraphics[width=.45\textwidth]{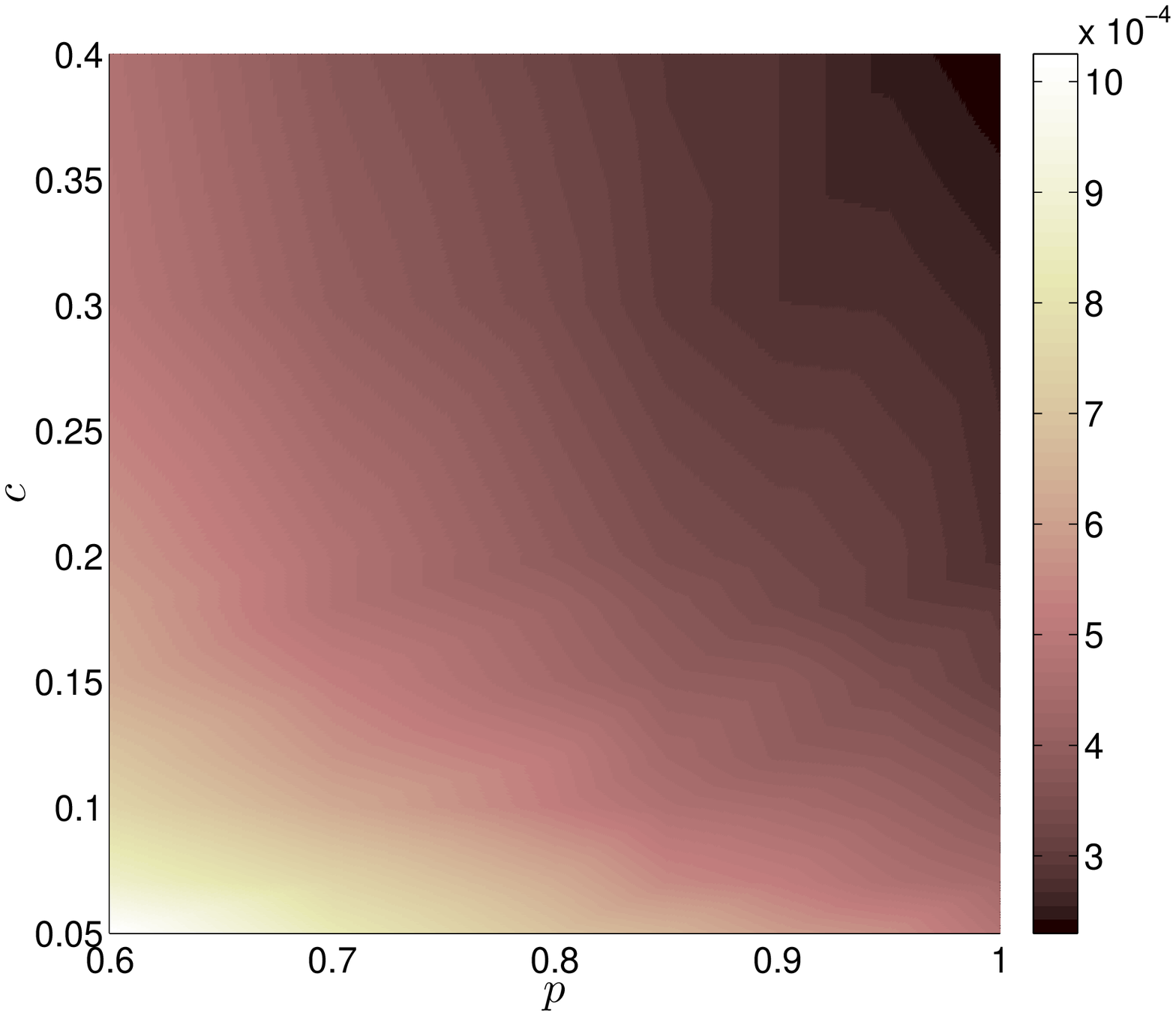}
   \includegraphics[width=.50\textwidth]{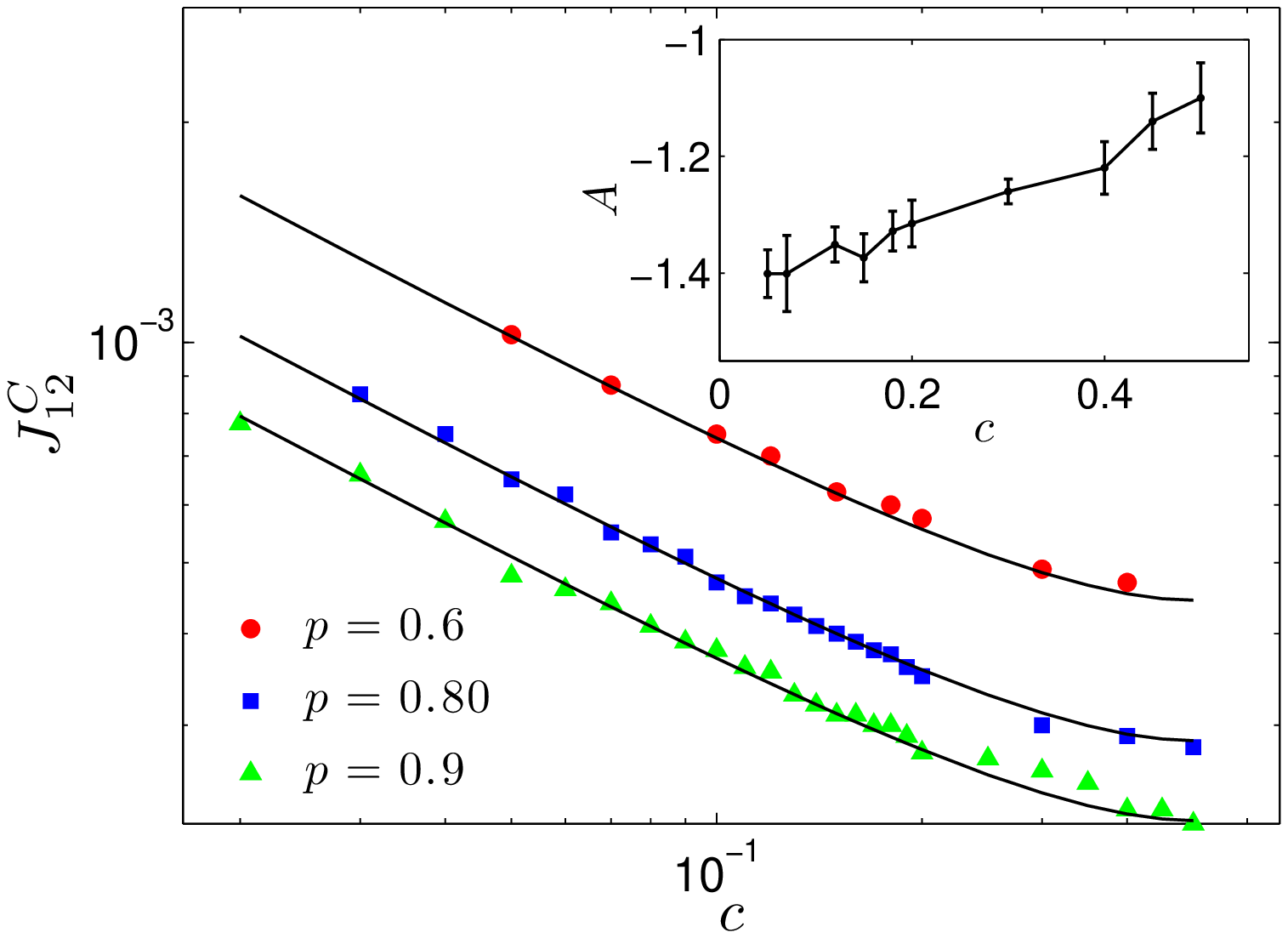}
  \caption{Left panel: Contour plot for measured values of $J_{12}^C$ as a function of $p$ and $c$. Right panel: Main figure: critical coupling $J_{12}^C$ as a function of $c$ for different values of link probability $p$, each represented with a different symbol, as shown by the legend; the continuous line represent the best fit according to Eq.~\ref{eq:c_dep}; Inset: Exponent $A$ (see Eq.~\ref{eq:alfa_dep}) obtained by fitting data for $J_{12}^C$ versus $p$, for different values of $c$. Both panels refer to a system of volume $N=6000$, and fixed internal couplings $J_{11}=J_{22}=10^{-4}$.}
  \label{fig:J12C_alpha_c}
  \label{fig:Fit}
\end{figure}

Let us now focus on the dependence of $J_{12}^C$ on the system parameters.
As shown in Fig.~\ref{fig:J12C_alpha_c}, $J_{12}^C$ grows as $c$ and/or $p$ get smaller, namely the more inhomogeneous are the two parties and/or the less connected is the graph, and the larger the interaction coupling necessary to induce a phase transition. This behavior is easy to see recalling that the expected number of  inter-group bonds is $N_1 N_2 p = N \alpha c (1-c)$, so that when the concentration of inter-group bonds is small, the coupling strength $J_{12}$ has to be larger in order to trigger a phase transition.

We can further deepen this point starting from Eq.~\ref{eq:contucci}: As $J_{12} \to J_{12}^C$, $M_1$ and $M_2$ are both close to zero, so that a first-order expansion leads to
\begin{equation}
  \left\{
    \begin{array}{rl}
      M_1 \approx J_{11}cM_1 + J_{12}^C (1-c)M_2, \nonumber \\
      M_2 \approx J_{12}^C cM_1 + J_{22}(1-c)M_2
    \end{array}
  \right.
\end{equation}
from which we get
\begin{equation}\label{eq:J_quad}
  (J_{12}^C)^2 \approx \frac{(1 - J_{11}c) [1-J_{22}(1-c)]}{(1-c)c},
\end{equation}
namely, recalling $J_{11}, J_{22} \ll 1$,
\begin{equation}\label{eq:c_dep}
  J_{12}^C \approx \frac{1}{\sqrt{(1-c)c}}.
\end{equation}

Let us now introduce the dilution. Our ansatz is that $J_{12}^C$ scales with $p$ according to a power law, giving the following overall expression
\begin{equation}\label{eq:alfa_dep}
J_{12}^C \approx \frac{p^{A(c)}}{\sqrt{c(1-c)}},
\end{equation}
where the exponent $A(c)$ depends on the inhomogeneity between the two populations; of course, when $p = 1$, the expression in Eq.~\ref{eq:c_dep} found for $K_N$ is naturally restored.

The fits performed on numerical data confirm the ansatz and show that the exponent $A(c)$ approaches the value $-1$ from below as $c$ goes to $0.5$ (see the inset of Fig.~\ref{fig:Fit}); otherwise stated, as the system gets more inhomogeneous, i.e. $c$ far from $1/2$, the role of the underlying topology, encoded by $p$, gets more important. Moreover, when $c$ is small, the dependence on $c$ displayed by $A$ is negligible, i.e. $A(c) \approx \tilde{A} \approx -1.4$ so that the functional form of $J_{12}^C$ can be factorized into $J_{12}^C \approx p^{\tilde{A}} / \sqrt{c(1-c)}$, which was used to fit data in Fig.~\ref{fig:Fit} (right panel, main figure).

\subsection{Critical exponents}
The power-law behaviors of Eqs.~\ref{eq:m} and ~\ref{eq:chi} are consistent with our numerical data and by means of fitting procedures we were able to estimate the critical exponents $\beta$ and $\gamma$, for different values of $c$ and $\alpha$.

\begin{figure}[ht]
  \centering
  \includegraphics[width=.90\textwidth]{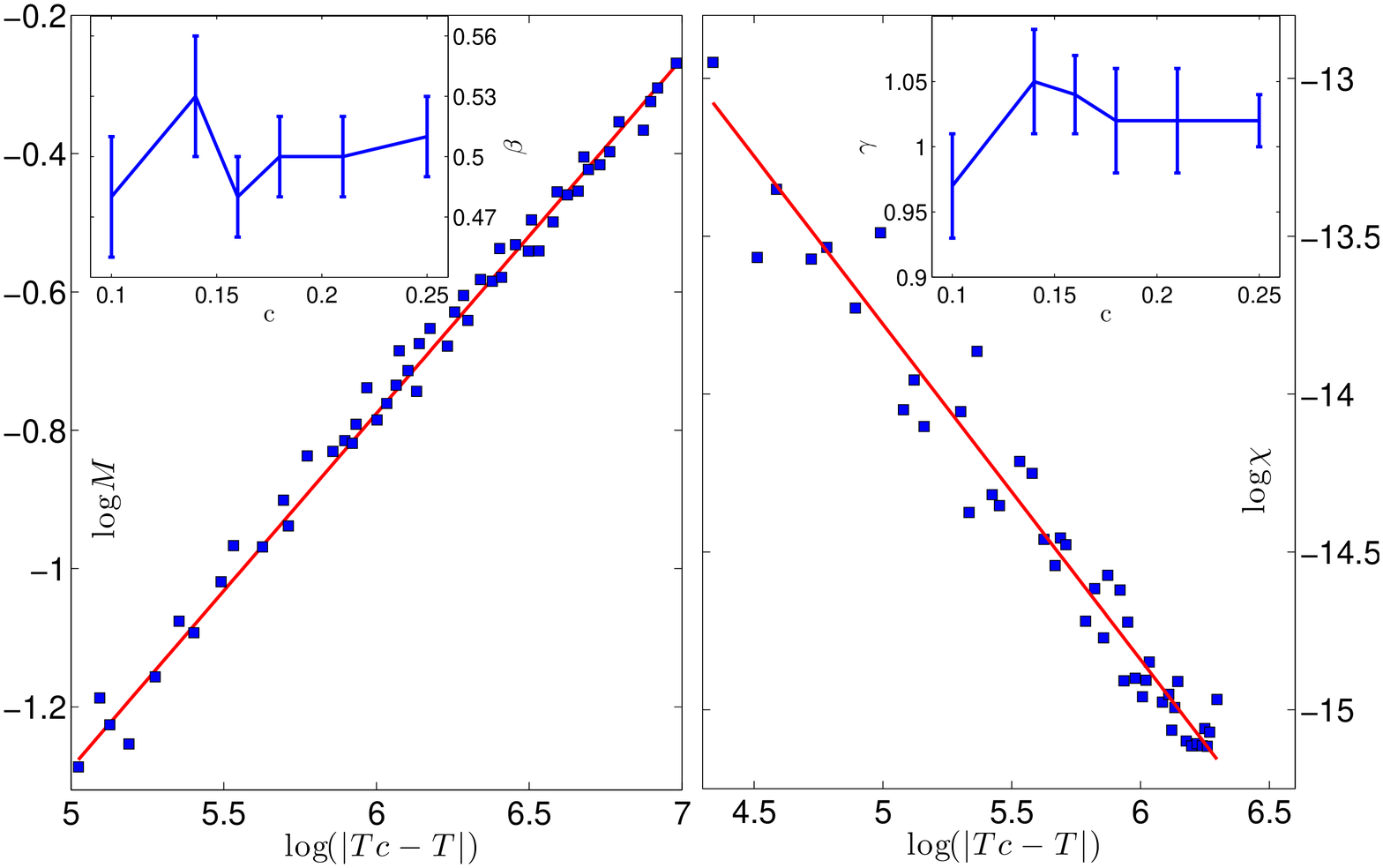}
  \caption{System made up of $N=6000$ nodes, with $J_{11}=J_{22}=10^{-4}$ and $p=0.80$. Main figures: $\log M$ (left panel) and $\log \chi$ (right panel) versus $\log \left| T_C-T \right|$ for the case $c=0.18$. Insets: critical exponent $\beta$ (left panel) and $\gamma$ (right panel) for different values of $c$.}
  \label{fig:expcritici}
\end{figure}

In Fig.~\ref{fig:expcritici} we show an example of fitting curve for $\beta$ and $\gamma$, respectively, found for a given realization of the substrate with $c=0.18$; values of the exponents obtained in this way for several configurations at fixed $c$ are then averaged in order to get the final estimate depicted in the inset as a function of concentrations.
Our estimates of the critical exponents are, within the error ($\sim 5 \%$), independent of $c$ and are consistent with the values known for the Curie-Weiss model and the diluted Ising ferromagnet \cite{agliari1}, i.e. $\beta=1/2$ and $\gamma=-1$. Similar results were obtained also for different values of connectivity ($\alpha > N/4$). Therefore, such a universality class exhibits a large degree of robustness: its properties are not affected by the (uncorrelated) dilution of the underlying structure and neither by a block-matrix coupling.

\subsection{Coupling rescaling}
The expression found for $J_{12}^C$ allows to rescale the curves of the magnetization: for fixed $J_{11} = J_{22} = J$, we write the magnetization of each population as a function of a rescaled coupling $\tilde{J}_{12}$ 
\begin{eqnarray}\label{eq:ansatz1}
M_1 = f_1(\tilde{J}_{12}, \alpha, c), \\
\label{eq:ansatz2}
M_2 = f_2(\tilde{J}_{12}, \alpha, c), 
\end{eqnarray}
where $f_1$ and $f_2$ are proper functions of the system parameters and 
\begin{equation} \label{eq:scaling}
\tilde{J}_{12} = (J_{12} - J_{12}^C) J^{B(c)}.
\end{equation}
Following the arguments and the results of Sec.~\ref{ssec:coupling}, we expect that a large concentration of inter-population links ($c=1/2$) gives rise to a less intensive dependence on $J$, i.e. $B$ is close to zero, and vice versa. In fact, for $c=1/2$, we can first consider the MF equations (Eq.~\ref{eq:solContucci}) and notice that, due to the symmetry, $M_1 = \tanh(J M_1/2 + J_{12} M_1/2) = \tanh((\tilde{J}_{12} + 2)M_1 /2)$, and analogously for $M_2$, where we used $J_{12}^C \approx 2 - J$, holding for $c=1/2$. Hence, $M_1$ and $M_2$ both depends on $\tilde{J}_{12}$ only, which yields $B(1/2)=0$. On the other hand, for the homogeneous case $c=0$ (or, symmetrically, $c=1$), the one-population diluted Ising model with constant coupling $J$ is recovered \cite{agliari1}. In Fig.~\ref{fig:robusto} the magnetization for $c=0.20$ and $p=0.80$ is plotted versus the rescaled coupling $\tilde{J}_{12}$ and we used $B(0.20)=1/3$. Analogous results hold for different concentrations $c$ as shown in Fig.~$4$, where the comparison with the MF theory (dashed curves) is also depicted. 
Notice that, for both diluted and complete graphs, when the system is more inhomogeneous, i.e $c$ is small, the scaling is less effective.

\begin{figure}[ht]
  \centering
  \includegraphics[width=.85\textwidth]{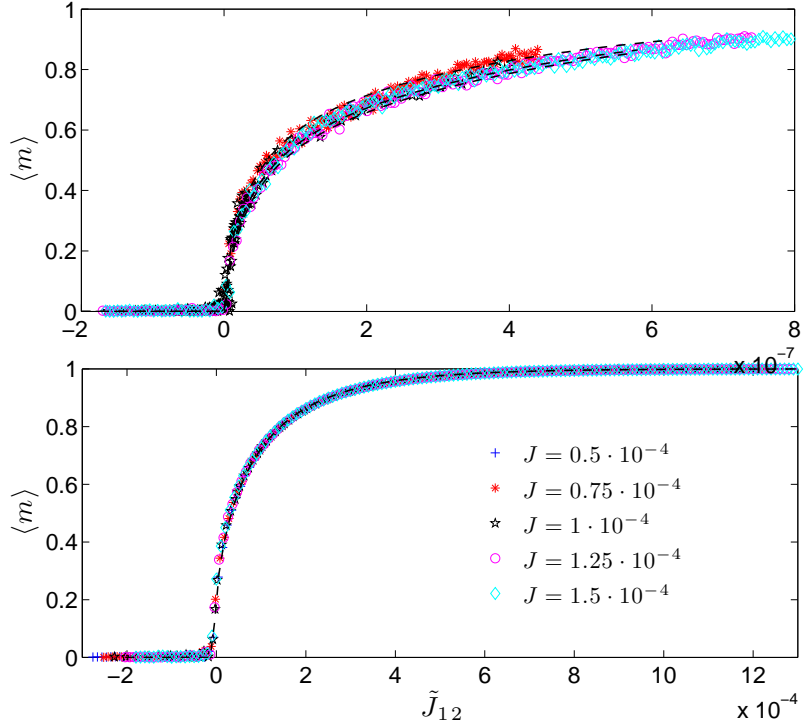} \label{fig:scala}
  \caption{Comparison between the MF solution from Eq.~\ref{eq:contucci} (dashed line) and numerical data for the diluted case, being $p =0.8$ and $c=0.1$ (upper panel) and $c=0.5$ (lower panel); different values of $J$ are considered, as shown in the legend.}
\end{figure}

\section{Conclusions} \label{sec:conclusions}
In this work we have analyzed the critical properties of the two-populations Ising model defined on Erd\"{o}s-R\'enyi random graphs.  This model features two degrees of inhomogeneity interplaying, namely topological dilution and  a four-blocks coupling matrix, which make the analytical treatment definitely awkward and, on the other hand, make the model rather rich also in view of possible applications. 
In general, a unitary and rigorous physical description of critical phenomena in disordered systems still lacks and the study of further models for which there is a general agreement in the behavior of the corresponding pure cases is very important. In this sense, the two-population Ising ferromagnet is a further suitable candidate for testing the above predictions that has not been previously investigated in the literature. 

The system is characterized by the topological parameter $\alpha = p N$, which represents the average number of links stemming from a node and by the number of nodes $N$ making up the system, which is divided into two subsets of cardinality $N_1$ and $N_2$, being $N_1+N_2=N$ and $c=N_1/N$; moreover the inter-group coupling is encoded by $J_{12}$, while the intra-group coupling by $J_{11}$ and $J_{22}$, respectively. We focused our attention on the case of both $J_{11}$ and $J_{22}$ small, while $J_{12}$ is tuned playing the role of a ``temperature''.  

We ran extensive Monte Carlo simulations and we measured magnetization and susceptibility for the whole system evidencing the occurring of a critical point at $J_{12}^C$, which has been shown to scale like $p^{A(c)} /\sqrt{c(1-c)}$; the exponent $A(c)$ approaches from below the value $-1$ as the system gets more homogeneous, namely as $c \to 1/2$. Therefore, the role of the underlying topology gets more important the larger the imbalance among the two populations.

We also measured the critical exponents $\beta$ and $\gamma$ showing that they are consistent with those pertaining to the Curie-Weiss model and to the diluted Ising model. This result highlights the robustness of the universality class, not only with respect the dilution of the underlying structure, but also with respect to the existence of two populations.

\section*{Acknowledgments}
The authors are grateful to P. Contucci, A. Barra and A. Schianchi for useful discussions and suggestions.

\end{document}